\documentclass[
 reprint,
 superscriptaddress,
 showpacs,
 nofootinbib,
 amsmath,amssymb,
 aps,
 prl
]{revtex4-1}

\usepackage{graphicx}
\usepackage{dcolumn}
\usepackage{bm}

\def \nobreakseq {\nobreak \hskip 0pt \hbox}

\newcommand{\subfigimg}[3][,]{%
  \setbox1=\hbox{\includegraphics[#1]{#3}}%
  \leavevmode\rlap{\usebox1}%
  \rlap{\hspace*{35pt}\raisebox{\dimexpr\ht1-3\baselineskip}{#2}}%
  \phantom{\usebox1}%
}

\begin{document}

\title{Search for neutrinoless quadruple-$\beta$ decay of $^{150}$Nd with the NEMO-3 detector}

\author{R.~Arnold}
\affiliation{IPHC, ULP, CNRS/IN2P3\nobreakseq{,} F-67037 Strasbourg, France}
\author{C.~Augier} 
\affiliation{LAL, Universit\'{e} Paris-Sud\nobreakseq{,} CNRS/IN2P3\nobreakseq{,} Universit\'{e} Paris-Saclay\nobreakseq{,} F-91405 Orsay\nobreakseq{,} France}
\author{A.S.~Barabash}
\affiliation{NRC ``Kurchatov Institute", ITEP, 117218 Moscow, Russia}
\author{A.~Basharina-Freshville} 
\affiliation{UCL, London WC1E 6BT\nobreakseq{,} United Kingdom}
\author{S.~Blondel} 
\affiliation{LAL, Universit\'{e} Paris-Sud\nobreakseq{,} CNRS/IN2P3\nobreakseq{,} Universit\'{e} Paris-Saclay\nobreakseq{,} F-91405 Orsay\nobreakseq{,} France}
\author{S.~Blot}
\affiliation{University of Manchester\nobreakseq{,} Manchester M13 9PL\nobreakseq{,}~United Kingdom}
\author{M.~Bongrand} 
\affiliation{LAL, Universit\'{e} Paris-Sud\nobreakseq{,} CNRS/IN2P3\nobreakseq{,} Universit\'{e} Paris-Saclay\nobreakseq{,} F-91405 Orsay\nobreakseq{,} France}
\author{D.~Boursette}
\affiliation{LAL, Universit\'{e} Paris-Sud\nobreakseq{,} CNRS/IN2P3\nobreakseq{,} Universit\'{e} Paris-Saclay\nobreakseq{,} F-91405 Orsay\nobreakseq{,} France}
\author{V.~Brudanin} 
\affiliation{JINR, 141980 Dubna, Russia}
\affiliation{National Research Nuclear University MEPhI, 115409 Moscow, Russia}
\author{J.~Busto} 
\affiliation{Aix Marseille Universit\'e\nobreakseq{,} CNRS\nobreakseq{,} CPPM\nobreakseq{,} F-13288 Marseille\nobreakseq{,} France}
\author{A.J.~Caffrey}
\affiliation{Idaho National Laboratory\nobreakseq{,} Idaho Falls, ID 83415, U.S.A.}
\author{S.~Calvez}
\affiliation{LAL, Universit\'{e} Paris-Sud\nobreakseq{,} CNRS/IN2P3\nobreakseq{,} Universit\'{e} Paris-Saclay\nobreakseq{,} F-91405 Orsay\nobreakseq{,} France}
\author{M.~Cascella} 
\affiliation{UCL, London WC1E 6BT\nobreakseq{,} United Kingdom}
\author{C.~Cerna} 
\affiliation{CENBG\nobreakseq{,} Universit\'e de Bordeaux\nobreakseq{,} CNRS/IN2P3\nobreakseq{,} F-33175 Gradignan\nobreakseq{,} France}
\author{J.P.~Cesar}
\affiliation{University of Texas at Austin\nobreakseq{,} Austin\nobreakseq{,} TX 78712\nobreakseq{,}~U.S.A.}
\author{A.~Chapon} 
\affiliation{LPC Caen\nobreakseq{,} ENSICAEN\nobreakseq{,} Universit\'e de Caen\nobreakseq{,} CNRS/IN2P3\nobreakseq{,} F-14050 Caen\nobreakseq{,} France}
\author{E.~Chauveau} 
\affiliation{University of Manchester\nobreakseq{,} Manchester M13 9PL\nobreakseq{,}~United Kingdom}
\author{A.~Chopra} 
\affiliation{UCL, London WC1E 6BT\nobreakseq{,} United Kingdom}
\author{L.~Dawson} 
\affiliation{UCL, London WC1E 6BT\nobreakseq{,} United Kingdom}
\author{D.~Duchesneau} 
\affiliation{LAPP, Universit\'e de Savoie\nobreakseq{,} CNRS/IN2P3\nobreakseq{,} F-74941 Annecy-le-Vieux\nobreakseq{,} France}
\author{D.~Durand} 
\affiliation{LPC Caen\nobreakseq{,} ENSICAEN\nobreakseq{,} Universit\'e de Caen\nobreakseq{,} CNRS/IN2P3\nobreakseq{,} F-14050 Caen\nobreakseq{,} France}
\author{V.~Egorov}
\affiliation{JINR, 141980 Dubna, Russia}
\author{G.~Eurin} 
\affiliation{LAL, Universit\'{e} Paris-Sud\nobreakseq{,} CNRS/IN2P3\nobreakseq{,} Universit\'{e} Paris-Saclay\nobreakseq{,} F-91405 Orsay\nobreakseq{,} France}
\affiliation{UCL, London WC1E 6BT\nobreakseq{,} United Kingdom}
\author{J.J.~Evans} 
\affiliation{University of Manchester\nobreakseq{,} Manchester M13 9PL\nobreakseq{,}~United Kingdom}
\author{L.~Fajt} 
\affiliation{Institute of Experimental and Applied Physics\nobreakseq{,} Czech Technical University in Prague\nobreakseq{,} CZ-12800 Prague\nobreakseq{,} Czech Republic}
\author{D.~Filosofov} 
\affiliation{JINR, 141980 Dubna, Russia}
\author{R.~Flack} 
\affiliation{UCL, London WC1E 6BT\nobreakseq{,} United Kingdom}
\author{X.~Garrido} 
\affiliation{LAL, Universit\'{e} Paris-Sud\nobreakseq{,} CNRS/IN2P3\nobreakseq{,} Universit\'{e} Paris-Saclay\nobreakseq{,} F-91405 Orsay\nobreakseq{,} France}
\author{H.~G\'omez} 
\affiliation{LAL, Universit\'{e} Paris-Sud\nobreakseq{,} CNRS/IN2P3\nobreakseq{,} Universit\'{e} Paris-Saclay\nobreakseq{,} F-91405 Orsay\nobreakseq{,} France}
\author{B.~Guillon} 
\affiliation{LPC Caen\nobreakseq{,} ENSICAEN\nobreakseq{,} Universit\'e de Caen\nobreakseq{,} CNRS/IN2P3\nobreakseq{,} F-14050 Caen\nobreakseq{,} France}
\author{P.~Guzowski} 
\affiliation{University of Manchester\nobreakseq{,} Manchester M13 9PL\nobreakseq{,}~United Kingdom}
\author{R.~Hod\'{a}k} 
\affiliation{Institute of Experimental and Applied Physics\nobreakseq{,} Czech Technical University in Prague\nobreakseq{,} CZ-12800 Prague\nobreakseq{,} Czech Republic}
\author{A.~Huber} 
\affiliation{CENBG\nobreakseq{,} Universit\'e de Bordeaux\nobreakseq{,} CNRS/IN2P3\nobreakseq{,} F-33175 Gradignan\nobreakseq{,} France}
\author{P.~Hubert} 
\affiliation{CENBG\nobreakseq{,} Universit\'e de Bordeaux\nobreakseq{,} CNRS/IN2P3\nobreakseq{,} F-33175 Gradignan\nobreakseq{,} France}
\author{C.~Hugon}
\affiliation{CENBG\nobreakseq{,} Universit\'e de Bordeaux\nobreakseq{,} CNRS/IN2P3\nobreakseq{,} F-33175 Gradignan\nobreakseq{,} France}
\author{S.~Jullian} 
\affiliation{LAL, Universit\'{e} Paris-Sud\nobreakseq{,} CNRS/IN2P3\nobreakseq{,} Universit\'{e} Paris-Saclay\nobreakseq{,} F-91405 Orsay\nobreakseq{,} France}
\author{A.~Klimenko} 
\affiliation{JINR, 141980 Dubna, Russia}
\author{O.~Kochetov} 
\affiliation{JINR, 141980 Dubna, Russia}
\author{S.I.~Konovalov} 
\affiliation{NRC ``Kurchatov Institute", ITEP, 117218 Moscow, Russia}
\author{V.~Kovalenko}
\affiliation{JINR, 141980 Dubna, Russia}
\author{D.~Lalanne} 
\affiliation{LAL, Universit\'{e} Paris-Sud\nobreakseq{,} CNRS/IN2P3\nobreakseq{,} Universit\'{e} Paris-Saclay\nobreakseq{,} F-91405 Orsay\nobreakseq{,} France}
\author{K.~Lang} 
\affiliation{University of Texas at Austin\nobreakseq{,} Austin\nobreakseq{,} TX 78712\nobreakseq{,}~U.S.A.}
\author{Y.~Lemi\`ere} 
\affiliation{LPC Caen\nobreakseq{,} ENSICAEN\nobreakseq{,} Universit\'e de Caen\nobreakseq{,} CNRS/IN2P3\nobreakseq{,} F-14050 Caen\nobreakseq{,} France}
\author{T.~Le~Noblet} 
\affiliation{LAPP, Universit\'e de Savoie\nobreakseq{,} CNRS/IN2P3\nobreakseq{,} F-74941 Annecy-le-Vieux\nobreakseq{,} France}
\author{Z.~Liptak} 
\affiliation{University of Texas at Austin\nobreakseq{,} Austin\nobreakseq{,} TX 78712\nobreakseq{,}~U.S.A.}
\author{X.~R.~Liu} 
\affiliation{UCL, London WC1E 6BT\nobreakseq{,} United Kingdom}
\author{P.~Loaiza} 
\affiliation{LAL, Universit\'{e} Paris-Sud\nobreakseq{,} CNRS/IN2P3\nobreakseq{,} Universit\'{e} Paris-Saclay\nobreakseq{,} F-91405 Orsay\nobreakseq{,} France}
\author{G.~Lutter} 
\affiliation{CENBG\nobreakseq{,} Universit\'e de Bordeaux\nobreakseq{,} CNRS/IN2P3\nobreakseq{,} F-33175 Gradignan\nobreakseq{,} France}
\author{M.~Macko}
\affiliation{FMFI,~Comenius~University\nobreakseq{,}~SK-842~48~Bratislava\nobreakseq{,}~Slovakia}
\affiliation{CENBG\nobreakseq{,} Universit\'e de Bordeaux\nobreakseq{,} CNRS/IN2P3\nobreakseq{,} F-33175 Gradignan\nobreakseq{,} France}
\author{C.~Macolino}
\affiliation{LAL, Universit\'{e} Paris-Sud\nobreakseq{,} CNRS/IN2P3\nobreakseq{,} Universit\'{e} Paris-Saclay\nobreakseq{,} F-91405 Orsay\nobreakseq{,} France}
\author{F.~Mamedov}
\affiliation{Institute of Experimental and Applied Physics\nobreakseq{,} Czech Technical University in Prague\nobreakseq{,} CZ-12800 Prague\nobreakseq{,} Czech Republic}
\author{C.~Marquet} 
\affiliation{CENBG\nobreakseq{,} Universit\'e de Bordeaux\nobreakseq{,} CNRS/IN2P3\nobreakseq{,} F-33175 Gradignan\nobreakseq{,} France}
\author{F.~Mauger} 
\affiliation{LPC Caen\nobreakseq{,} ENSICAEN\nobreakseq{,} Universit\'e de Caen\nobreakseq{,} CNRS/IN2P3\nobreakseq{,} F-14050 Caen\nobreakseq{,} France}
\author{B.~Morgan} 
\affiliation{University of Warwick\nobreakseq{,} Coventry CV4 7AL\nobreakseq{,} United Kingdom}
\author{J.~Mott} 
\affiliation{UCL, London WC1E 6BT\nobreakseq{,} United Kingdom}
\author{I.~Nemchenok} 
\affiliation{JINR, 141980 Dubna, Russia}
\author{M.~Nomachi} 
\affiliation{Osaka University\nobreakseq{,} 1-1 Machikaneyama Toyonaka\nobreakseq{,} Osaka 560-0043\nobreakseq{,} Japan}
\author{F.~Nova} 
\affiliation{University of Texas at Austin\nobreakseq{,} Austin\nobreakseq{,} TX 78712\nobreakseq{,}~U.S.A.}
\author{F.~Nowacki} 
\affiliation{IPHC, ULP, CNRS/IN2P3\nobreakseq{,} F-67037 Strasbourg, France}
\author{H.~Ohsumi} 
\affiliation{Saga University\nobreakseq{,} Saga 840-8502\nobreakseq{,} Japan}
\author{C.~Patrick} 
\affiliation{UCL, London WC1E 6BT\nobreakseq{,} United Kingdom}
\author{R.B.~Pahlka}
\affiliation{University of Texas at Austin\nobreakseq{,} Austin\nobreakseq{,} TX 78712\nobreakseq{,}~U.S.A.}
\author{F.~Perrot} 
\affiliation{CENBG\nobreakseq{,} Universit\'e de Bordeaux\nobreakseq{,} CNRS/IN2P3\nobreakseq{,} F-33175 Gradignan\nobreakseq{,} France}
\author{F.~Piquemal} 
\affiliation{CENBG\nobreakseq{,} Universit\'e de Bordeaux\nobreakseq{,} CNRS/IN2P3\nobreakseq{,} F-33175 Gradignan\nobreakseq{,} France}
\affiliation{Laboratoire Souterrain de Modane\nobreakseq{,} F-73500 Modane\nobreakseq{,} France}
\author{P.~Povinec}
\affiliation{FMFI,~Comenius~University\nobreakseq{,}~SK-842~48~Bratislava\nobreakseq{,}~Slovakia}
\author{P.~P\v{r}idal} 
\affiliation{Institute of Experimental and Applied Physics\nobreakseq{,} Czech Technical University in Prague\nobreakseq{,} CZ-12800 Prague\nobreakseq{,} Czech Republic}
\author{Y.A.~Ramachers} 
\affiliation{University of Warwick\nobreakseq{,} Coventry CV4 7AL\nobreakseq{,} United Kingdom}
\author{A.~Remoto}
\affiliation{LAPP, Universit\'e de Savoie\nobreakseq{,} CNRS/IN2P3\nobreakseq{,} F-74941 Annecy-le-Vieux\nobreakseq{,} France}
\author{J.L.~Reyss} 
\affiliation{LSCE\nobreakseq{,} CNRS\nobreakseq{,} F-91190 Gif-sur-Yvette\nobreakseq{,} France}
\author{C.L.~Riddle} 
\affiliation{Idaho National Laboratory\nobreakseq{,} Idaho Falls, ID 83415, U.S.A.}
\author{E.~Rukhadze} 
\affiliation{Institute of Experimental and Applied Physics\nobreakseq{,} Czech Technical University in Prague\nobreakseq{,} CZ-12800 Prague\nobreakseq{,} Czech Republic}
\author{R.~Saakyan} 
\affiliation{UCL, London WC1E 6BT\nobreakseq{,} United Kingdom}
\author{R.~Salazar} 
\affiliation{University of Texas at Austin\nobreakseq{,} Austin\nobreakseq{,} TX 78712\nobreakseq{,}~U.S.A.}
\author{X.~Sarazin} 
\affiliation{LAL, Universit\'{e} Paris-Sud\nobreakseq{,} CNRS/IN2P3\nobreakseq{,} Universit\'{e} Paris-Saclay\nobreakseq{,} F-91405 Orsay\nobreakseq{,} France}
\author{Yu.~Shitov} 
\affiliation{JINR, 141980 Dubna, Russia}
\affiliation{Imperial College London\nobreakseq{,} London SW7 2AZ\nobreakseq{,} United Kingdom}
\author{L.~Simard} 
\affiliation{LAL, Universit\'{e} Paris-Sud\nobreakseq{,} CNRS/IN2P3\nobreakseq{,} Universit\'{e} Paris-Saclay\nobreakseq{,} F-91405 Orsay\nobreakseq{,} France}
\affiliation{Institut Universitaire de France\nobreakseq{,} F-75005 Paris\nobreakseq{,} France}
\author{F.~\v{S}imkovic} 
\affiliation{FMFI,~Comenius~University\nobreakseq{,}~SK-842~48~Bratislava\nobreakseq{,}~Slovakia}
\author{A.~Smetana}
\affiliation{Institute of Experimental and Applied Physics\nobreakseq{,} Czech Technical University in Prague\nobreakseq{,} CZ-12800 Prague\nobreakseq{,} Czech Republic}
\author{K.~Smolek} 
\affiliation{Institute of Experimental and Applied Physics\nobreakseq{,} Czech Technical University in Prague\nobreakseq{,} CZ-12800 Prague\nobreakseq{,} Czech Republic}
\author{A.~Smolnikov} 
\affiliation{JINR, 141980 Dubna, Russia}
\author{S.~S\"oldner-Rembold}
\affiliation{University of Manchester\nobreakseq{,} Manchester M13 9PL\nobreakseq{,}~United Kingdom}
\author{B.~Soul\'e}
\affiliation{CENBG\nobreakseq{,} Universit\'e de Bordeaux\nobreakseq{,} CNRS/IN2P3\nobreakseq{,} F-33175 Gradignan\nobreakseq{,} France}
\author{D.~\v{S}tef{\'a}nik} 
\affiliation{FMFI,~Comenius~University\nobreakseq{,}~SK-842~48~Bratislava\nobreakseq{,}~Slovakia}
\author{I.~\v{S}tekl} 
\affiliation{Institute of Experimental and Applied Physics\nobreakseq{,} Czech Technical University in Prague\nobreakseq{,} CZ-12800 Prague\nobreakseq{,} Czech Republic}
\author{J.~Suhonen} 
\affiliation{Jyv\"askyl\"a University\nobreakseq{,} FIN-40351 Jyv\"askyl\"a\nobreakseq{,} Finland}
\author{C.S.~Sutton} 
\affiliation{MHC\nobreakseq{,} South Hadley\nobreakseq{,} Massachusetts 01075\nobreakseq{,} U.S.A.}
\author{G.~Szklarz}
\affiliation{LAL, Universit\'{e} Paris-Sud\nobreakseq{,} CNRS/IN2P3\nobreakseq{,} Universit\'{e} Paris-Saclay\nobreakseq{,} F-91405 Orsay\nobreakseq{,} France}
\author{J.~Thomas} 
\affiliation{UCL, London WC1E 6BT\nobreakseq{,} United Kingdom}
\author{V.~Timkin} 
\affiliation{JINR, 141980 Dubna, Russia}
\author{S.~Torre} 
\affiliation{UCL, London WC1E 6BT\nobreakseq{,} United Kingdom}
\author{Vl.I.~Tretyak} 
\affiliation{Institute for Nuclear Research\nobreakseq{,} 03028\nobreakseq{,} Kyiv\nobreakseq{,} Ukraine}
\author{V.I.~Tretyak}
\affiliation{JINR, 141980 Dubna, Russia}
\author{V.I.~Umatov} 
\affiliation{NRC ``Kurchatov Institute", ITEP, 117218 Moscow, Russia}
\author{I.~Vanushin} 
\affiliation{NRC ``Kurchatov Institute", ITEP, 117218 Moscow, Russia}
\author{C.~Vilela} 
\affiliation{UCL, London WC1E 6BT\nobreakseq{,} United Kingdom}
\author{V.~Vorobel} 
\affiliation{Charles University in Prague\nobreakseq{,} Faculty of Mathematics and Physics\nobreakseq{,} CZ-12116 Prague\nobreakseq{,} Czech Republic}
\author{D.~Waters} 
\affiliation{UCL, London WC1E 6BT\nobreakseq{,} United Kingdom}
\author{F.~Xie} 
\affiliation{UCL, London WC1E 6BT\nobreakseq{,} United Kingdom}
\author{A.~\v{Z}ukauskas}
\affiliation{Charles University in Prague\nobreakseq{,} Faculty of Mathematics and Physics\nobreakseq{,} CZ-12116 Prague\nobreakseq{,} Czech Republic}
\collaboration{NEMO-3 Collaboration}
\noaffiliation

\date{\today}

\begin{abstract}
We report the results of a first experimental search for lepton number violation by four units in the neutrinoless quadruple-$\beta$ decay of $^{150}$Nd using a total exposure of $0.19$~kg$\cdot$y recorded with the NEMO-3 detector at the Modane Underground Laboratory (LSM). 
  We find no evidence of this decay and set lower limits on the half-life in the range $T_{1/2}>(1.1\text{--}3.2)\times10^{21}$~y at the $90\%$ CL, depending on the model used for the kinematic distributions of the emitted electrons. 
\end{abstract}


\maketitle
In the standard model (SM) of particle physics, leptons are assigned a lepton number of $+1$ and anti-leptons are assigned $-1$. All experimental observations thus far
are consistent with the assumption that the total lepton number $L$ is conserved in particle interactions~\cite{pdg}.
However, since this is not due to a fundamental symmetry, there is no reason to assume that $L$ is generally conserved in theories beyond the SM. 

Lepton-number violating processes could be directly linked to the possible Majorana nature of neutrinos. If Majorana mass terms are added to the SM Lagrangian, processes appear that violate $L$ by two units ($\Delta L=2$)~\cite{valle}. Searches for $\Delta L=2$ processes such as neutrinoless double-$\beta$ ($0\nu2\beta$) decay have therefore been the focus of many experiments~\cite{double-beta-searches}.

In this letter, we present a first search for processes with $\Delta L=4$, which are allowed even if neutrinos are Dirac fermions and $\Delta L=2$ processes are forbidden~\cite{quad-theory}.
Models with $\Delta L=4$ have some power in explaining naturally small Dirac masses of neutrinos~\cite{dirac-mass} and could mediate leptogenesis~\cite{leptogenesis}. The models have also been linked with dark matter candidates~\cite{DM} and with \emph{CP} violation in the lepton sector~\cite{CP}. Processes with $\Delta L=4$ could also be probed at the Large Hadron Collider (LHC), for example
in the pair production and decay of triplet-Higgs states to four identical charged leptons~\cite{LHC}.

An experimental signature of some models with $\Delta L=4$ would be the neutrinoless quadruple-$\beta$ ($0\nu 4\beta$) decay of a nucleus,
$(A,Z) \to (A,Z+4) + 4 e^-$,
where four electrons are emitted with a total kinetic energy equal to the energy $Q_{4\beta}$ of the nuclear transition. The $0\nu4\beta$ half-life is expected to depend strongly on the unknown mass scale $\Lambda_\textrm{NP}$ of the new $\Delta L=4$ phenomena~\cite{quad-theory}.

The search for $0\nu 4\beta$ decay is experimentally challenging, since only three long-lived isotopes can undergo this decay, $^{136}$Xe ($Q_{4\beta}=0.079$~MeV~\cite{qvals}), $^{96}$Zr ($Q_{4\beta}=0.642$~MeV), and $^{150}$Nd, which has the highest $Q_{4\beta}$ value of $2.084$~MeV. The NEMO-3 detector contained two of these isotopes, $^{96}$Zr and $^{150}$Nd. The $^{96}$Zr decay has too low a $Q_{4\beta}$ value to be detected with high enough efficiency in NEMO-3 since low-energy electrons would be absorbed in the source. It could instead be studied using geochemical methods~\cite{zr96-geochem}. The value of $Q_{4\beta}$ of the decay $^{150}$Nd$\to^{150}$Gd, however, is sufficiently large for four electrons to be observable in the NEMO-3 detector.

We search for $0\nu 4\beta$ decay by exploiting the unique ability of the NEMO-3 experiment to reconstruct the kinematics of each final-state electron.
In the absence of a more complete theoretical treatment of the kinematics of the decay~\cite{heeck}, we test four models of the electron energy distributions, labeled uniform, symmetric, semi-symmetric, and anti-symmetric. This choice is designed to cover a wide range of models and used to demonstrate that the final result is largely model-independent. 

The uniform model has all four electron kinetic energies $T_i$ distributed uniformly on the simplex $T_1+T_2+T_3+T_4=Q_{4\beta}$ with each kinetic energy $T_i>0$.
The decay rates $\mathrm{d}N$ for the other three models are distributed according to the differential phase space given by
\begin{eqnarray}
\lefteqn{
\frac{\mathrm{d^4}N}{\prod_{i=1}^4\mathrm{d}T_i} 
  \propto } \\ & &   A_m \delta\left(Q_{4\beta}-\sum_{i=1}^4 T_i\right)\cdot 
    \prod_{i=1}^4 (T_i+m_e)p_i F(T_i, Z) , \nonumber
\label{eqn:kinem}
\end{eqnarray}
which is an extension of the $0\nu2\beta$-decay phase space~\cite{supernemo}.
Here, $i$ labels the electrons, $m_e$ the electron mass, $p_i = \sqrt{T_i(T_i+2m_e)}$, and $A_m$ is a model-dependent factor. The Fermi function 
$F(T, Z) \propto p^{2s-2} e^{\pi u} \left| \Gamma(s+iu)\right|^2$
describes the Coulomb attraction between the electrons and the daughter nucleus with atomic number $Z$. In this  function, $s=\sqrt{1-(\alpha Z)^{2}}$, $u=\alpha Z (T+m_e) / p$, $\Gamma$ is the gamma function, and $\alpha$ is the fine structure constant.
The three different phase space distributions differ by the factors $A_m$ that depend on the energy asymmetry of electron pairs. For the symmetric distribution $A_m=\mathcal{S}\{1\times1\}$, the semi-symmetric distribution has $A_m=\mathcal{S}\{1\times(T_k-T_l)^2\}$, and for the anti-symmetric distribution $A_m=\mathcal{S}\{(T_i-T_j)^2\times(T_k-T_l)^2\}$, where $\mathcal{S}\{\cdots\}$ is a sum over symmetric interchange of labels $i,j,k,l$ of the four electrons.
In all models, each electron angular distribution is generated isotropically.

\begin{figure}[htpb]
\includegraphics[width=1.1\columnwidth]{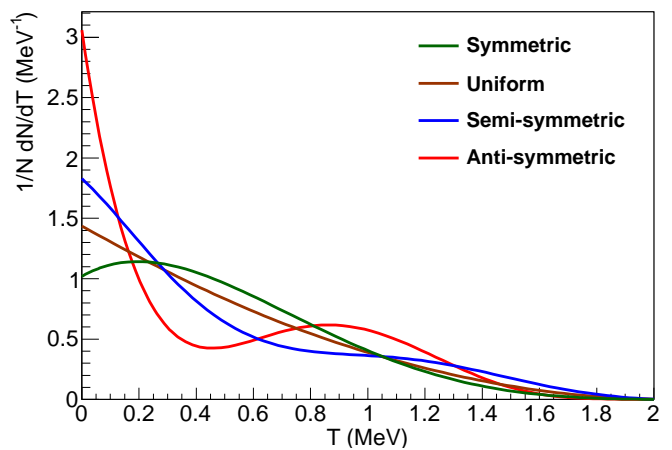}
\caption{Normalized distribution of the individual electron kinetic energies $T$ in each $0\nu 4\beta$ decay for the four kinematic models.}
\label{fig:true_energies}
\end{figure}

Since electrons produced in the NEMO-3 source foil must have a minimum energy of $\approx 250$~keV to fall into the acceptance, the efficiency is smaller for models producing more low-energy electrons. We show the electron kinetic energy distributions for the four kinematic models in Fig.~\ref{fig:true_energies}.

We perform the search with the NEMO-3 detector on data collected between $2003$ and $2011$ using $36.6$~g of enriched $^{150}$Nd source, with
a live time of $5.25$~y. The detector is optimized to search for
$0\nu 2\beta$ decays by reconstructing the full decay topology.
It is cylindrical in shape, with the cylinder axis oriented vertically, a height of $3$~m and a
diameter of $5$~m, and is divided into $20$ sectors of equal size.
Thin foils with a thickness of $40\text{--}60$ mg/cm$^2$
contain $7$ different isotopes.
The Nd foil has a height of $2.34$~m and a width of $6.5$~cm.
The foils are located between two concentric tracking  chambers composed of $6180$ drift cells operating in Geiger mode. Surrounding the tracking chambers on all sides are calorimeter walls composed of $1940$ scintillator blocks coupled to low-activity photomultipliers that provide
timing and energy measurements. The calorimeter energy
resolution is $(14.1\text{--}17.7)\%$ (FWHM) at an electron energy
of $1$~MeV. A vertically oriented magnetic field of $\approx 25$~G allows discrimination between electrons and positrons. Detailed descriptions of the experiment and data sets are given in Refs.~\cite{detector,mo}.

In Ref.~\cite{nd150-2beta}, we describe a measurement of the two-neutrino double-$\beta$ ($2\nu2\beta$) decay of $^{150}$Nd, and provide details of the background model and measured activities that are used in this analysis. The backgrounds are categorized as internal (within the source foil, including contamination of $^{208}$Tl and $^{214}$Bi), external to the foil (electrons and photons produced in or outside of the detector components), radon diffusion that can deposit background isotopes on the surface of the detector components, and also internal contamination in the source foils neighboring the Nd foil, which can have a falsely reconstructed vertex in the Nd foil. 

Internal conversions, M{\o}ller and Compton scattering are sources of additional electrons in single-$\beta$ or double-$\beta$ decays that can mimic four-electron final states.
The largest contribution to the background is $2\nu2\beta$ decay of $^{150}$Nd 
to the ground state (g.s.) of $^{150}$Sm with a half-life of $T_{1/2}=9.34\times10^{18}$~y~\cite{nd150-2beta}. An additional background source not considered in Ref.~\cite{nd150-2beta} is the double-$\beta$ decay of $^{150}$Nd to the $0^+_1$ excited state of $^{150}$Sm~\cite{nd150-new-excited}, for which we use a half-life of $T_{1/2}=1.33\times10^{20}$~y~\cite{nd150-excited} in the simulation.

The selection requires candidate decays that produce three or four tracks originating in the foil. If there are three tracks, all three must be matched to calorimeter hits, which is the signature of a reconstructed electron candidate, while the fourth $\beta$ electron is assumed to be absorbed in the foil ($3e$ topology). We further distinguish two topologies in the four-track final state, 
where either all four tracks are associated with calorimeter hits ($4e$ topology) or 
one of the tracks has no calorimeter hit ($3e1t$ topology).

\begin{figure}[htpb]
\subfigimg[width=0.9\linewidth]{{\bf(a)}}{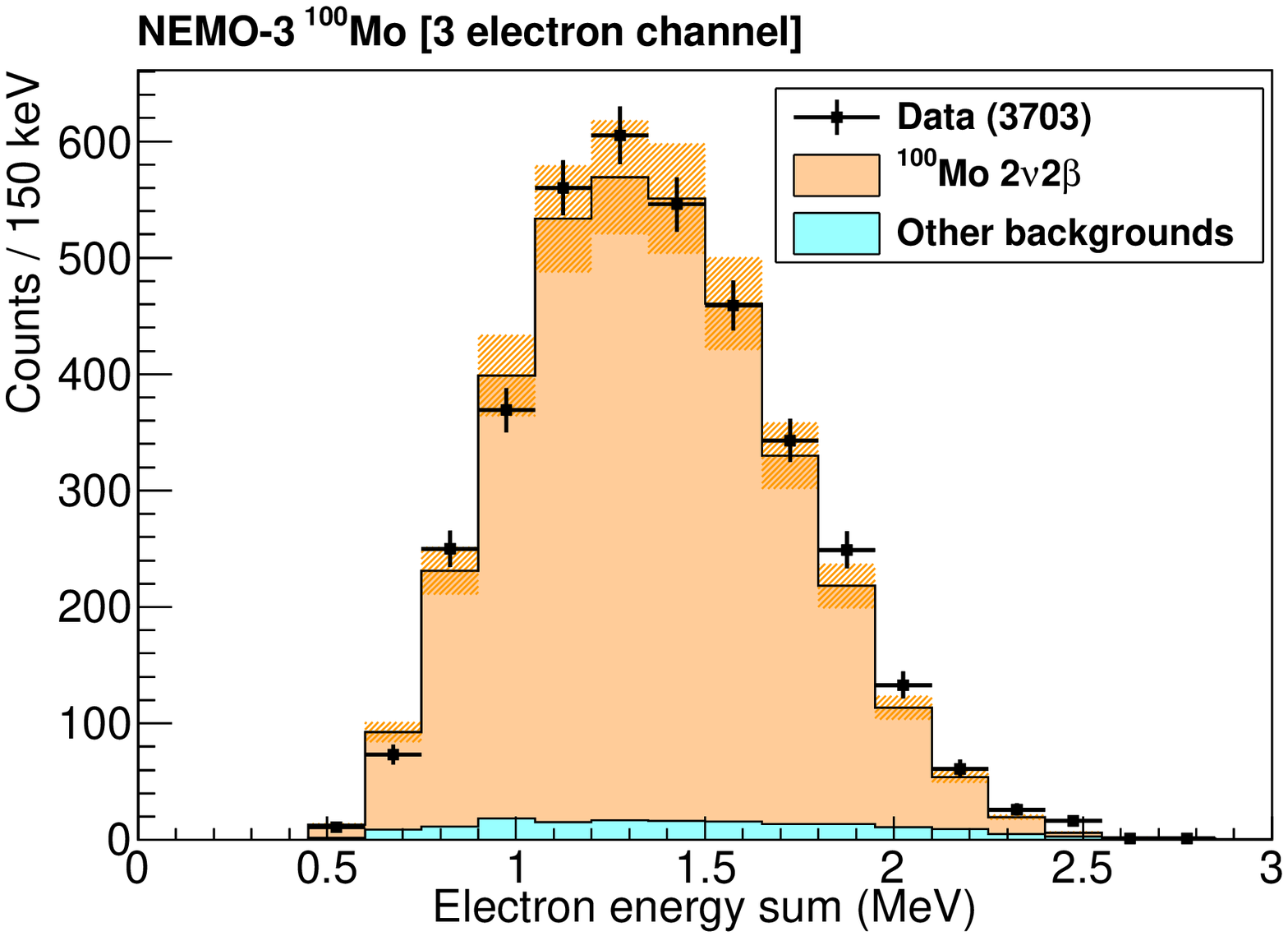}
\subfigimg[width=0.9\linewidth]{{\bf(b)}}{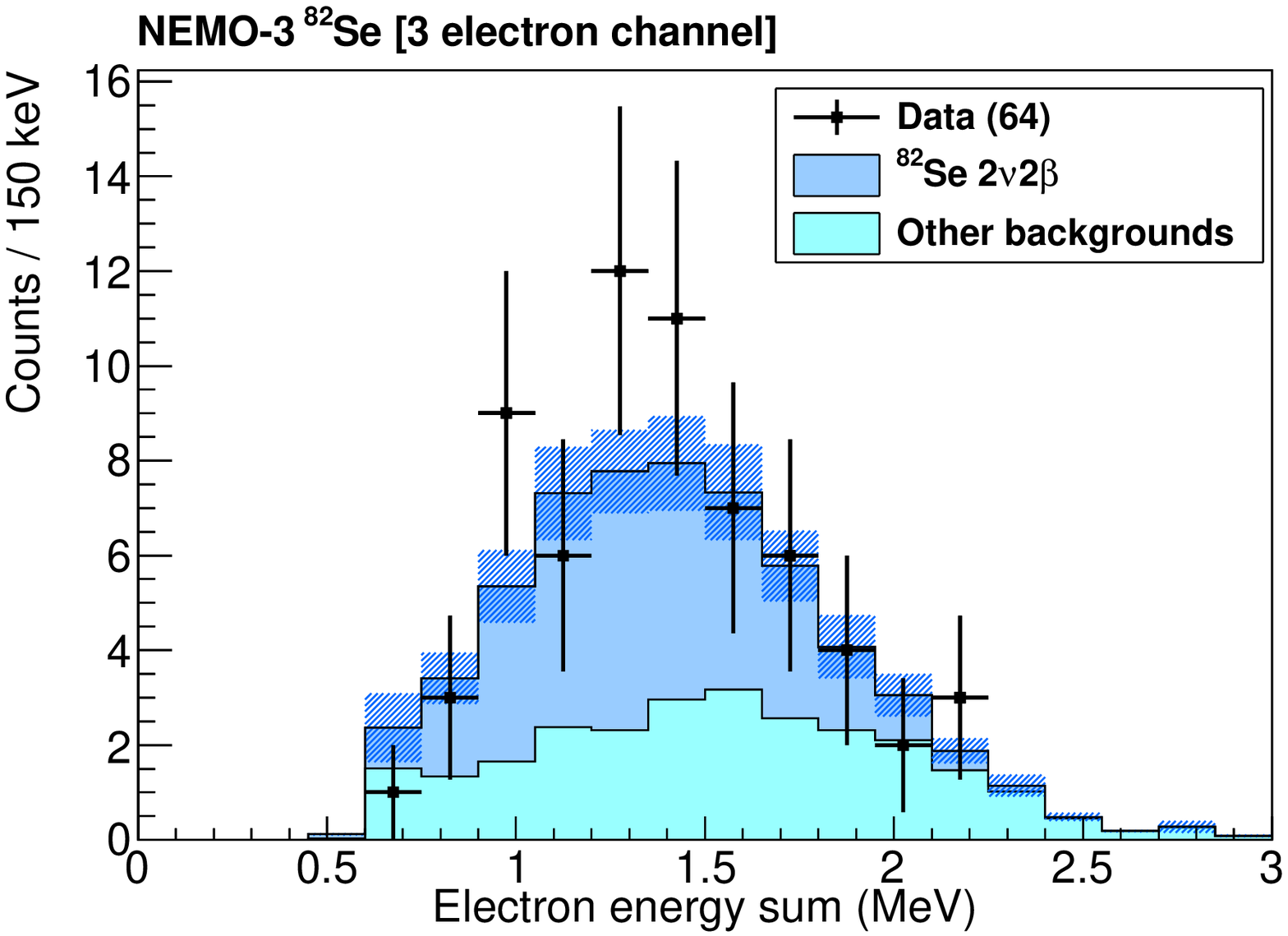}
  \caption{Energy-sum distributions for 
three-electron events originating in the source foils of (a) $^{100}$Mo and (b) $^{82}$Se, which cannot undergo $0\nu4\beta$ decay. The hashed areas represent the systematic uncertainty on the reconstruction efficiency.}
\label{fig:sidebands}
\end{figure}

An additional set of selections is applied to all topologies to ensure events are well reconstructed
and to reject instrumental backgrounds. Decay vertices in regions of high activity in the foil corresponding to localized contaminations from $^{234m}$Pa and $^{207}$Bi (hot spots) are rejected.
The locations of these hot spots have been determined in Ref.~\cite{nd150-2beta}.
Events where more than one electron track is associated with the same calorimeter hit are removed. 
The energy of each associated calorimeter hit must be $>150$~keV. Events in the $4e$ topology with one associated calorimeter hit below $150$~keV are treated as $3e1t$ candidates.
The vertical component of the distance between the intersection points of the tracks with the foil must be $<8$~cm. We apply no requirement in the horizontal direction, since the foil has a width of $6.5$~cm.
For each event, the track lengths, calorimeter hit times and energies, along with their uncertainties, 
are used to construct two $\chi^2$ values assuming all tracks originate in the foil (internal
hypothesis) or one track originates outside the foil and scatters in the foil producing secondary 
tracks (external). 
The probabilities of the internal hypothesis must be $>0.1\%$ and of the external hypothesis $<4\%$. 
Finally, events with unassociated calorimeter hits with energies $>150$~keV in time with the electron candidates are rejected, since this would indicate that photons were emitted in the decay. 

\begin{table}[tb]
\begin{tabular}{lcccc}
  \hline\hline
Topology & Symmetric & Uniform & Semi-symm. & Anti-symm. \\
\hline
4e    & 0.20 & 0.13 & 0.04 & 0.01 \\
3e    & 3.55 & 3.11 & 2.39 & 1.67 \\
3e1t  & 0.86 & 0.64 & 0.30 & 0.13 \\
\hline
Total & 4.61 & 3.88 & 2.73 & 1.81 \\
  \hline\hline
\end{tabular}
  \caption{Signal efficiencies (in $\%$) of the four kinematic models for the three topologies.}
\label{tab:effs}
\end{table}

For the $3e1t$ topology only, we require that there are no delayed hits with times
up to $700$~$\mu$s near the decay vertex or the track end points, caused by an $\alpha$ decay of the $^{214}$Po daughter of $^{214}$Bi $\beta$ decays~\cite{nd150-2beta}. These decays can occur on the surface of the tracker wires with the $\beta$ electron scattering in the foil producing secondaries. The $\beta$ electron in this type of decay would have no associated calorimeter hit.

\begin{figure}[htpb]
\includegraphics[width=0.7\linewidth]{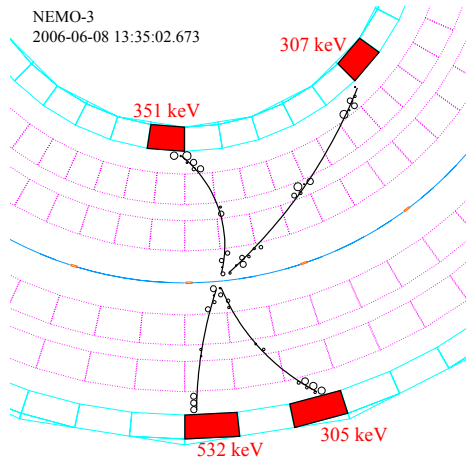}
\caption{Display of a decay with four reconstructed electrons in NEMO-3 data, originating in the $^{100}$Mo source foil, in the horizontal plane.}
\label{fig:evdisp}
\end{figure}

To validate the background model, the selection is applied to the foils containing the isotopes $^{100}$Mo and $^{82}$Se, which are expected to contain no $0\nu4\beta$ signal.
The energy-sum distributions for the $3e$ topology, which have higher statistics, are shown in Fig.~\ref{fig:sidebands}. We observe no events in the $4e$ topology in the $^{82}$Se foil, where $0.05\pm0.01$ are expected.  We observe two $4e$-candidates
in the $^{100}$Mo foil, with an expectation of $2.3\pm 0.5$ events, of which $2.0\pm0.4$ are due to $2\nu2\beta$ decays followed by double M{\o}ller scattering.  A display of one of these two data events is shown in Fig.~\ref{fig:evdisp}.

The total efficiencies for signal decays are shown in Tab.~\ref{tab:effs} and range from $1.81\%$ to $4.61\%$ depending on kinematic model. The expected background yields are given in Tab.~\ref{tab:backgrounds} for the energy range $1.2 \le \Sigma E \le 2.0$~MeV, where $\Sigma E$ is the electron energy sum, obtained by summing over the calorimeter hits for all reconstructed electrons.
All activities and systematic uncertainties, except for the $2\nu2\beta$ $0^+_1$ process, are taken from Ref.~\cite{nd150-2beta}.
  
\begin{table}[htbp]
  \begin{tabular}{lccc}
    \hline\hline
    Origin & 4e $[\times10^{-2}]$ & 3e & 3e1t $[\times10^{-2}]$ \\
    \hline
    $^{150}$Nd $2\nu2\beta$ (g.s.)    & $2.08\pm0.57$ & $9.43\pm0.84$ & $8.98\pm0.92$ \\
    $^{150}$Nd $2\nu2\beta$ ($0^+_1$) & $0.85\pm0.36$ & $2.39\pm0.63$ & $3.98\pm1.07$ \\
    $^{208}$Tl internal               & $0.74\pm0.15$ & $1.28\pm0.21$ & $5.37\pm1.21$ \\
    $^{214}$Bi internal               & $0.19\pm0.07$ & $0.74\pm0.18$ & $1.08\pm0.30$ \\
    Other internals                   &               & $0.82\pm0.11$ & $1.01\pm0.51$ \\
    Neighboring foils                 &               & $1.61\pm0.45$ & $1.95\pm1.91$ \\
    Radon                             &               & $0.43\pm0.15$ &               \\
    Externals                         &               & $0.12\pm0.09$ & $6.50\pm4.12$ \\
    \hline
    Total                             & $3.86\pm0.74$ & $16.8\pm1.7$  & $28.9\pm5.4$  \\
    \hline\hline
  \end{tabular}
  \caption{Expected number of background events for an exposure of $36.6$~g$\times5.25$~y in the $^{150}$Nd source foil in the range $1.2 \le \Sigma E \le 2.0$~MeV for the three topologies, with
  their total systematic uncertainties.}
  \label{tab:backgrounds}
\end{table}

\begin{figure}[htpb]
\subfigimg[width=0.9\linewidth]{\bf(a)}{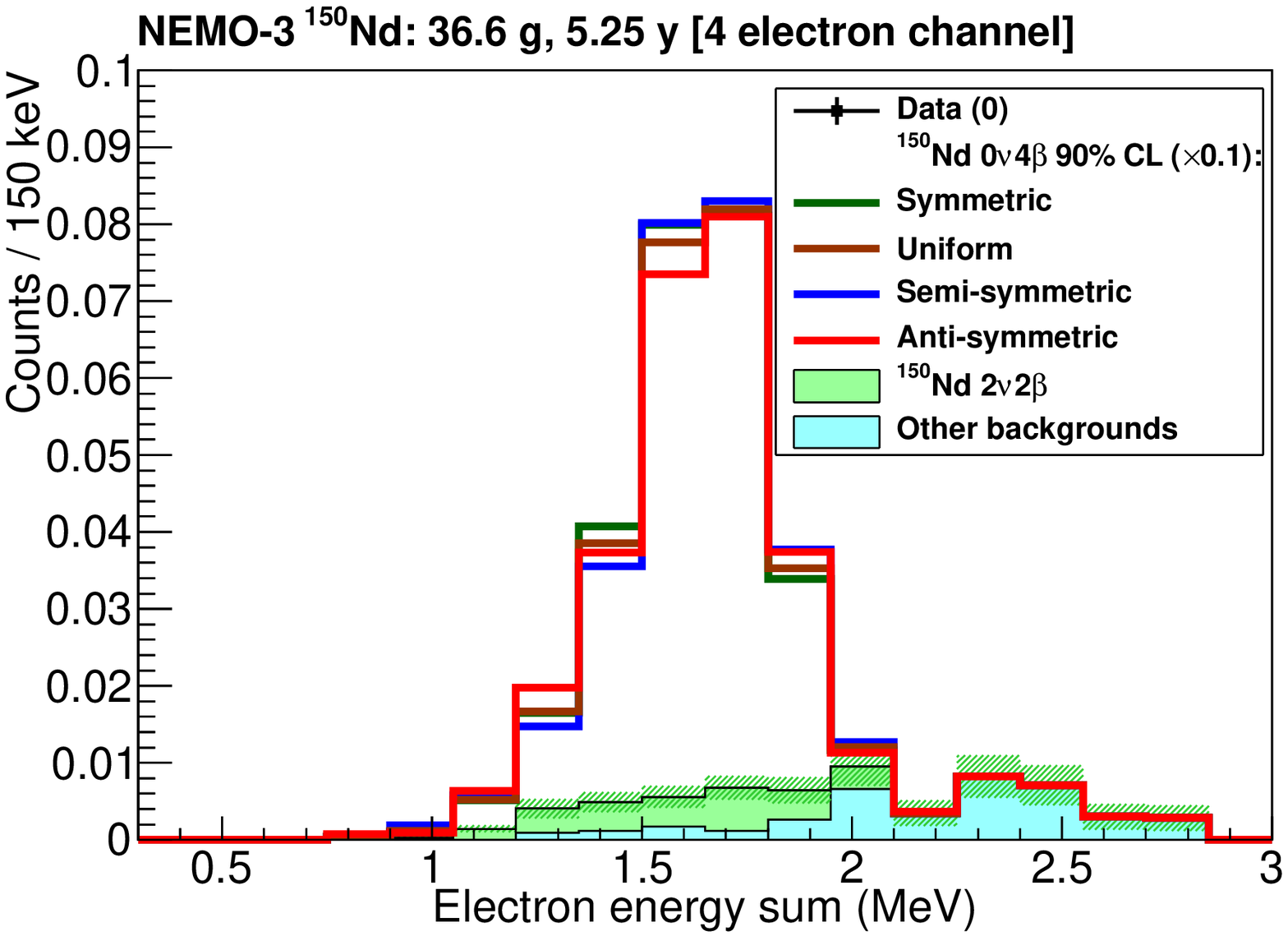}
\subfigimg[width=0.9\linewidth]{\bf(b)}{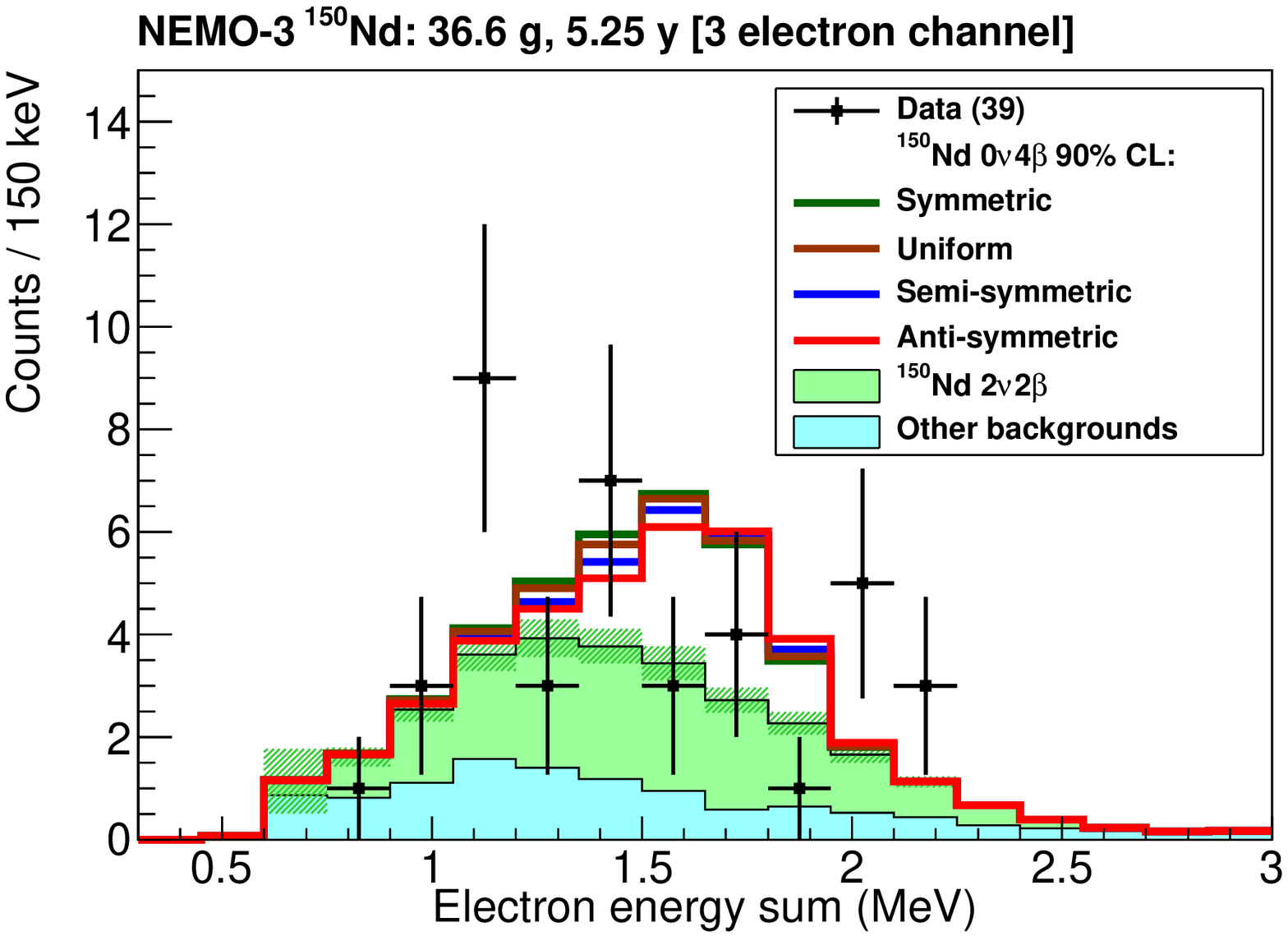}
\subfigimg[width=0.9\linewidth]{\bf(c)}{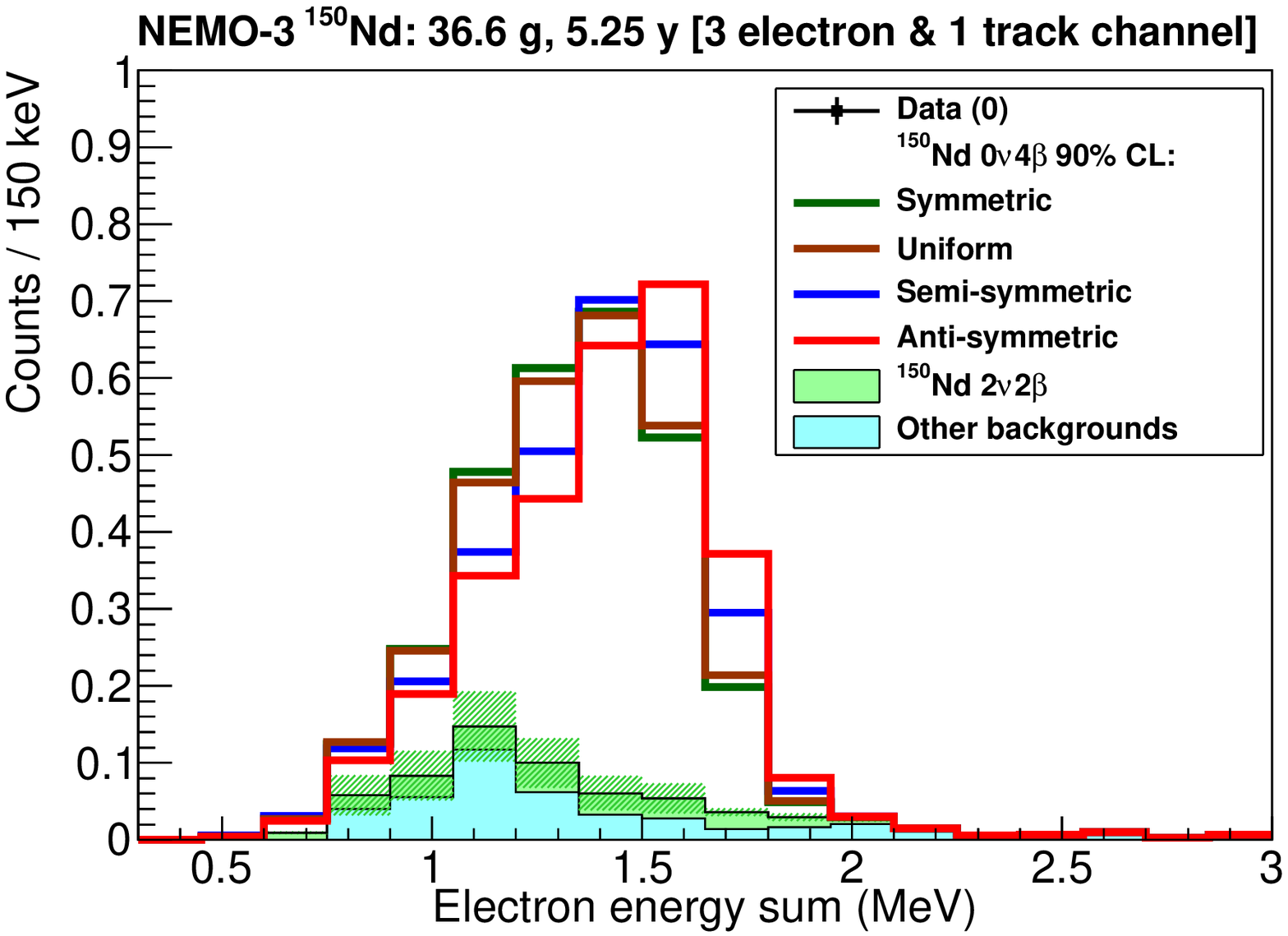}
  \caption{Energy sum distributions for (a) $4e$, (b) $3e$, and (c) $3e1t$ events in the $^{150}$Nd foil for data, the expected background and signal. The hashed areas represent the uncertainty on the background model. The $0\nu 4\beta$ signal distributions are normalized to
  the $90\%$ CL limits, with an additional scaling factor of $0.1$ applied to the expected $4e$ signal distributions for better visibility.}
\label{fig:ene_specs}
\end{figure}

The distributions of the electron energy-sum for events originating from the Nd foil are shown in Fig.~\ref{fig:ene_specs}. The energies of the signal distributions are lower than  $Q_{4\beta}=2.084$~MeV due to electron energy losses in the source foil. In addition, only three of the electrons have an associated calorimeter energy measurement for the $3e1t$ candidate events. The distributions show that there are no large differences between the shapes for the different kinematic models.

We observe no candidate events in the $4e$ and $3e1t$ topologies, with expected background rates of $0.04\pm0.01$ and $0.29 \pm 0.05$ events, respectively. There is also no significant excess of data in the $3e$ topology, with 22 observed events in the range $1.2 \le \Sigma E \le 2.0$~MeV, compared to $16.8 \pm 1.7$ expected background events.

\begin{table}[htbp]
\begin{tabular}{lccc}
\hline\hline
 Source & 4e & 3e & 3e1t \\
 \hline
 Reconstruction efficiency ($\epsilon_{2e}$) & $\pm5.5\%$ & $\pm5.5\%$ & $\pm5.5\%$ \\
 Reconstruction efficiency ($\epsilon_{3e}$) & $\pm 8.5\%$& $\pm 8.5\%$& $\pm 8.5\%$\\
 Energy scale & $\pm12.1\%$ & $\pm4.4\%$ & $\pm8.5\%$ \\
 Angular distribution & $\pm5.7\%$ & $\pm1.9\%$ & $\pm4.5\%$ \\
\hline\hline
\end{tabular}
\caption{Systematic uncertainties on the signal normalization for the three topologies.}
\label{tab:sigsyst}
\end{table}
We consider several sources of systematic uncertainty. The systematic uncertainties on the background model given in Tab.~\ref{tab:backgrounds} are the same as used in Ref.~\cite{nd150-2beta}, apart from the $25\%$ uncertainty on the half-life of the $2\nu 2\beta$ $0^+_1$ excited state decay~\cite{nd150-excited}.
The uncertainties of the signal efficiency are given in Tab.~\ref{tab:sigsyst}.
The uncertainty on the reconstruction efficiency is determined using the $^{100}$Mo data. It is broken down into two independent components, one based on a two-electron efficiency ($\epsilon_{2e}$) uncertainty of $5.5\%$ and the second on a three-electron
($\epsilon_{3e}$) uncertainty of $8.5\%$. The first value is obtained
by comparing the independently measured activity of a $^{207}$Bi calibration source with the in-situ measurements. This uncertainty can only be determined for decays with a maximum of two electrons in the final state.
The three-electron uncertainty ($\epsilon_{3e}$) of $8.5\%$ is obtained by comparing the normalization of the $3e$ selection in the simulation and data for the $^{100}$Mo foils. This additional uncertainty is taken to be correlated between signal and background, and assumed to be the same size in the $4e$ and $3e1t$ topologies.
A variation of $2\%$ on the energy scale for all electrons is applied in the simulation to cover uncertainties on the $Q_{4\beta}$ value, the calorimetric energy reconstruction of $0.2\%$~\cite{mo}, and uncertainties on the simulated energy loss in the foil.
The uncertainty due to the assumption of an isotropic angular distribution
of the electrons from the $0\nu 4\beta$ decay is derived from the variation of the reconstruction efficiency as a function of the generated angles between electron pairs. 

For the $4e$ and $3e1t$ topologies, where no candidate events
are observed, we set limits using a single bin for each topology, as for a counting experiment. For the $3e$ topology, we use the binned distribution of Fig.~\ref{fig:ene_specs}(b).
Limits at the $90\%$ CL are calculated using the modified-frequentist {\it CL}$_s$ method~\cite{cls}, which includes the systematic uncertainties with Gaussian priors.

The observed and expected half-life limits $T_{1/2}^{0\nu 4\beta}$ are shown in Tab.~\ref{tab:limits}. We obtain the best sensitivity in the $3e$ topology, due to the much higher signal efficiency compared to the $4e$ topology.
The combined lower limit at the $90\%$~CL on the $0\nu 4\beta$ half-life is  $3.2\times10^{21}$~y, with a sensitivity, given by the median expected limit, of $3.7\times10^{21}$~y, assuming a symmetric energy distribution. The combined
limits lie in the range $(1.1\mbox{--}3.2)\times10^{21}$~y for the different models. This result represents the first search for neutrinoless quadruple-$\beta$ decay in any isotope, and the first search for lepton-number violation by 4 units. 

\begin{table}[htbp]
\begin{tabular}{lcccccccc}
  \hline
  \hline
 & \multicolumn{2}{c}{Symmetric} & \multicolumn{2}{c}{~Uniform~~~} & \multicolumn{2}{c}{Semi-symm.} & \multicolumn{2}{c}{Anti-symm.} \\
 & obs & exp & obs & exp & obs & exp & obs & exp \\
\hline
4e         & 0.5 & 0.3 & 0.3 & 0.2 & 0.1 & 0.1 & 0.03 & 0.02 \\
3e         & 1.6 & 2.4 & 1.5 & 2.1 & 1.2 & 1.7 & 0.9 & 1.2 \\
3e1t       & 2.0 & 1.9 & 1.5 & 1.4 & 0.7 & 0.6 & 0.3 & 0.3 \\ 
Combined   & 3.2 & 3.7 & 2.6 & 3.0 & 1.7 & 2.0 & 1.1 & 1.3 \\
\hline\hline
\end{tabular}
\caption{Observed and median expected lower limits at the $90\%$ CL on the $0\nu4\beta$ half-life (in units of $10^{21}$~y) for the four signal models. Systematic uncertainties are taken into account with Gaussian priors.}
\label{tab:limits}
\end{table}

To improve on this limit in the future using the NEMO-3 technique would not only require more exposure, but also an optimization of the foil density and thickness which causes the main loss of efficiency for low-energy electrons and increases background from M{\o}ller scattering. Even with reduced isotope mass, a thinner foil should increase sensitivity. 

Since our search strategy is largely model-independent, this first limit on
$0\nu 4\beta$ decay can provide valuable constraints on new-physics models.
The authors of Ref.~\cite{quad-theory} estimate for their particular model the ratio $R$ of the $0\nu4\beta$ half-life to the $2\nu2\beta$ half-life to be $R\approx 10^{46} (\Lambda_\textrm{NP}/\textrm{TeV})^4$.
For $T_{1/2}^{0\nu 4\beta} > 1.1\times10^{21}~$years, this translates to a limit of $R> 120$.  For $0\nu 4\beta$ processes to be observable at the current experimental sensitivities, significant enhancement factors are therefore required.  
This result thus motivates further theoretical and experimental studies of $\Delta L=4$ processes in nuclear decays and
at colliders.  

The  authors  would  like  to  thank  the  staff  of  the
Modane Underground Laboratory for their technical assistance in operating the detector. We thank Werner Rodejohann for useful discussions.  
We acknowledge support by the funding agencies of the Czech Republic, the
National  Center  for  Scientific  Research/National  Institute of Nuclear and Particle Physics (France), the Russian Foundation for Basic Research (Russia), the Slovak Research and Development Agency, the Science
and  Technology  Facilities  Council and the Royal Society (United  Kingdom),
and the National Science Foundation (United States).

\end{document}